\documentclass[preprint,aps,floatfix,superscriptaddress,pre]{revtex4}
\usepackage{amsmath, amsthm, amssymb, graphicx}
\usepackage{color}
\usepackage{multirow}
\usepackage{ulem}

\input epsf.sty
\begin{document}

\title{Coarsening with non-trivial in-domain dynamics: correlations and interface fluctuations}
\author{Barton L. Brown}
\affiliation{Department of Physics, Virginia Tech, Blacksburg, VA 24061-0435, USA}
\affiliation{Center for Soft Matter and Biological Physics, Virginia Tech, Blacksburg, VA 24061-0435, USA}
\author{Michel Pleimling}
\affiliation{Department of Physics, Virginia Tech, Blacksburg, VA 24061-0435, USA}
\affiliation{Center for Soft Matter and Biological Physics, Virginia Tech, Blacksburg, VA 24061-0435, USA}
\affiliation{Academy of Integrated Science, Virginia Tech, Blacksburg, VA 24061-0405, USA}
\date{\today}

\begin{abstract}
Using numerical simulations we investigate the space-time properties of a system in which spirals emerge within coarsening
domains, thus giving rise to non-trivial internal dynamics. 
Initially proposed in the context of population dynamics, the studied six-species model
exhibits growing domains composed of three species in a rock-paper-scissors relationship. 
Through the investigation of different quantities, such as space-time correlations and the derived characteristic length,
autocorrelation, density of empty sites, and interface width, we demonstrate that the non-trivial dynamics inside
the domains affects the coarsening process as well as the properties of the interfaces separating different domains.
Domain growth, aging, and interface fluctuations are shown to be governed by exponents whose values differ from those
expected in systems with curvature driven coarsening.
\end{abstract}

\maketitle

\section{Introduction}

Systems out of equilibrium are often characterized by rich space-time patterns \cite{Cross93}, as for example in coarsening processes. Coarsening domains
are encountered in a variety of situations, ranging from magnetic systems quenched deep inside the ordered phase \cite{Bray94,Henkel10} to competing bacterial
colonies \cite{Szolnoki14} and social systems with opinion dynamics \cite{Castellano09}. Curvature driven phase ordering \cite{Puri09}
is a relaxation process ubiquitous in nature where the typical domain length increases with the square root of time. A well studied
case is provided by the two-dimensional Ising model quenched to temperatures below the critical point \cite{Bray94}.
Coarsening processes in more complex systems sometimes yield a much slower growth of the domains. For example, in 
disordered ferromagnets \cite{Rao93,Park10,Corberi12,Park12,Corberi13,Mandal14}, in systems composed of
elastic lines moving in disordered media \cite{Kolton05,Noh09,Iguain09} or in systems dominated by dynamical constraints 
\cite{Evans98,Lahiri97,Brown15} one observes domains that increase logarithmically with time.

In standard situations domain coarsening is characterized by two different time scales: a short time scale due to the 
microscopic degrees of freedom and a long time scale due to the motion of the domain walls. Consider as an example the
two-dimensional Ising model. Inside the ordered domains spins behave essentially like in the equilibrium steady state, with
some spins changing sign due to thermal noise. Spins remain in this quasi-equilibrium state as long as no domain wall crosses
through the region that contains the spins. It follows that spins deep inside an ordered region exhibit the trivial dynamics
of an equilibrium system.

The question we explore in this paper is whether and, if so, to what extent non-trivial dynamics inside a domain changes
the properties during coarsening and relaxation  processes. We address this through a study of many-species
models that have originally been proposed in the context of population dynamics involving predators and preys \cite{Roman13}.

Recent studies have shown that models used to describe predator-prey systems can display intriguing emerging phenomena when considering
a spatial setting and/or stochastic effects (see \cite{Szabo07} for a review of some early results). 
Much work has been devoted to cyclic cases as for example the three-species
cyclic game \cite{Frey10} or the corresponding game with four species where each species 
is preying on one other species while being at the same time the prey of another species \cite{Szabo04,
Case10,Durney11,Durney12,Roman12,Dobrinevski12,Intoy13,Guisoni13}. Whereas some earlier papers have 
considered spatial and stochastic effects in systems with a larger number of species \cite{Frachebourg96,Frachebourg96b,Frachebourg98,
Szabo01,Szabo01b,He05,Szabo05,Szabo07b,Szabo07c,Perc07,Szabo08,Szabo08b}, it is only in the last few years that systematic
theoretical studies of more complicated food networks with five or more species have become available 
\cite{Avelino12a,Avelino12b,Roman13,Knebel13,Kang13,Vukov13,Cheng14,Mowlaei14,Avelino14a,Avelino14b,Roman16,Kang16,
Labavic16,Avelino16}. One of the intriguing results of these studies has been the discovery of a rich variety of
space-time patterns, including spirals where each wavefront is formed by a single species, fuzzy spirals due to the
mixing of different species inside the waves, coarsening domains where every domain is formed by an alliance of mutually
neutral species as well as coarsening processes where inside every domain spirals are formed, thus yielding non-trivial dynamics
inside the coarsening domains \cite{Roman13,Mowlaei14,Roman16,Brown17}. In most cases a complete characterization of the spatio-temporal
properties has not yet been achieved.

In the following we aim to elucidate the space-time properties of the simplest system with non-trivial dynamics within the 
growing domains, namely a six-species model where in each domain three species undergo an effective cyclic rock-paper-scissors game \cite{Frey10}.
Our goal is to gain a rather complete picture of the relaxation processes in this system through a systematic study of various 
space and time-dependent quantities that allow us to capture many properties of the domains and the interfaces separating them.
We compare our results with those obtained from a modified version of the model that does not exhibit spirals within the domains as well
as with those from a model that exhibits coarsening due to the competition of only two species. Our results reveal
that the large-scale structures (i.e. spirals) formed inside the domains strongly impact domain formation, aging processes, as well
as interface fluctuations, yielding sets of exponents that differ from those expected for curvature-driven coarsening.

The paper is organized in the following way. In the next section we introduce our six-species model that is characterized
by the formation of spirals within coarsening domains when starting from a fully disordered initial state. We also discuss 
a variation of the six-species model that does not exhibit spirals as well as a system with only
two species that also undergoes coarsening.
In Section III we present a numerical investigation of our system. The study of a variety of
quantities (space-time correlations and derived correlation lengths, autocorrelation, density of empty sites, and 
interface fluctuations) yields a rather comprehensive picture of the relaxation processes in our systems.
Comparing results obtained from the different
models allows us to gain an understanding of how non-trivial dynamics within domains can change the spatio-temporal
properties of a coarsening process. 
In Section IV we discuss our results and conclude.

\section{Six-species model with coarsening and spirals}
The six-species model at the center of our study is a member of a broader family of May-Leonard type predator-prey models 
with symmetric interactions \cite{Roman13}. Using the notation proposed in previous work \cite{Roman13,Mowlaei14,Labavic16},
the general $(N,r)$ game consists of $N$ species, each preying on $r$ other species in a cyclic way. Playing these
(or related) games on a two-dimensional lattice yields a surprisingly rich variety of space-time patterns
\cite{Avelino12a,Avelino12b,Roman13,Mowlaei14,Avelino14a,Avelino14b,Roman16,Labavic16,Avelino16}, 
for example, coarsening domains composed of spiral structures or even spirals
nested within larger spirals \cite{Brown17}.

\begin{figure} [h]
\includegraphics[width=0.40\columnwidth,clip=true]{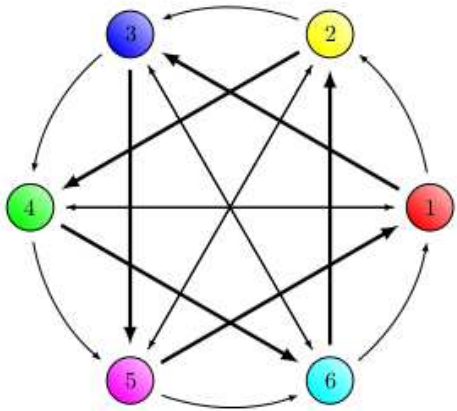}
\caption{\label{fig1} (Color online) Interaction diagram for the $(6,3)$ game. The arrows connect predators with their preys.
On a two-dimensional lattice two teams of cyclically interacting species each form their own domains, see Fig. \ref{fig2}. The bold 
arrows indicate the two teams of three species that emerge from this interaction scheme.
}
\end{figure}

Our main focus will be on the May-Leonard version of the $(6,3)$ game with the interaction network shown in Fig. \ref{fig1}. We consider a 
two-dimensional lattice where species interactions are limited to nearest neighbors. The possible
interactions can be summarized in the form of reactions taking place between neighboring sites:
\begin{equation}
\begin{split}
s_{i} + s_{j} &\xrightarrow[]{\kappa} s_{i} + \emptyset \\
s_{i} + \emptyset &\xrightarrow[]{\kappa} s_{i} + s_{i} \\
s_{i} + X &\xrightarrow[]{\sigma} X+s_{i} 
\end{split}
\label{eq:1}
\end{equation}
where $s_i$, $i = 1, \cdots, 6$, denotes an individual of the $i$th species. $\emptyset$ indicates an empty site,
whereas $X$ can be an individual from any species or an empty site. The first reaction describes a predation
event where with rate $\kappa$ an individual of species $j$, which is a prey of species $i$, is removed from
the lattice. The second reaction describes reproduction where with rate $\kappa$ an individual of species $i$ 
creates an offspring on an empty neighboring site. The mobility of the individuals can take place in two ways, summarized in the 
third reaction given in Eq. (\ref{eq:1}): individuals on neighboring sites can swap places with rate $\sigma$ or 
an individual can jump to an empty neighboring site with the same rate $\sigma$. We normalize rates such that
$\kappa + \sigma =  1$. The results presented in this paper have all been obtained for $\kappa = \sigma = 0.5$.

In our agent-based simulations we allow for at most one individual at each site. This is different from a recent
study \cite{Labavic16} where a variation of the $(6,3)$ game was investigated in two space dimensions with multiple
occupancy of a site and only on-site reactions. For every attempt at an update we randomly select a site before
randomly selecting one of the four nearest neighbors. The selected neighboring site is then updated using the
reaction scheme (\ref{eq:1}). One unit of time corresponds to $V$ proposed updates where $V$ is the total number
of sites in the system.

\begin{figure}
\minipage{0.30\textwidth}
  \centering
  \includegraphics[width=\linewidth]{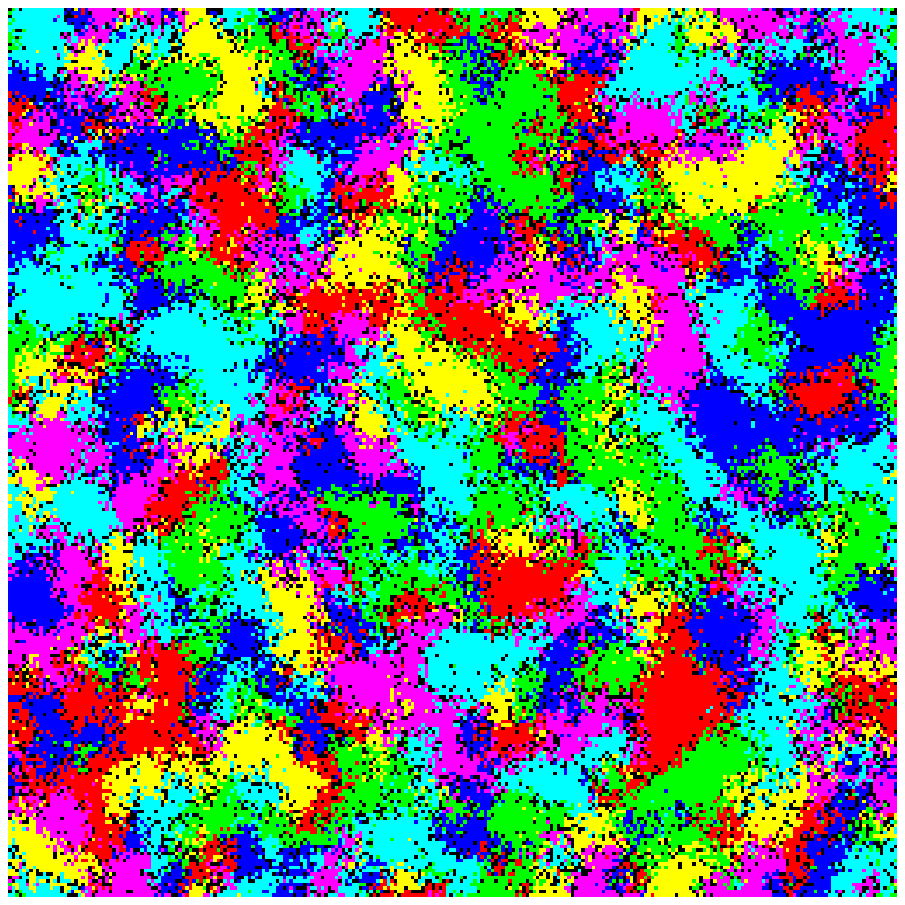}\\
  t=100
\endminipage\hfill
\minipage{0.30\textwidth}
  \centering
  \includegraphics[width=\linewidth]{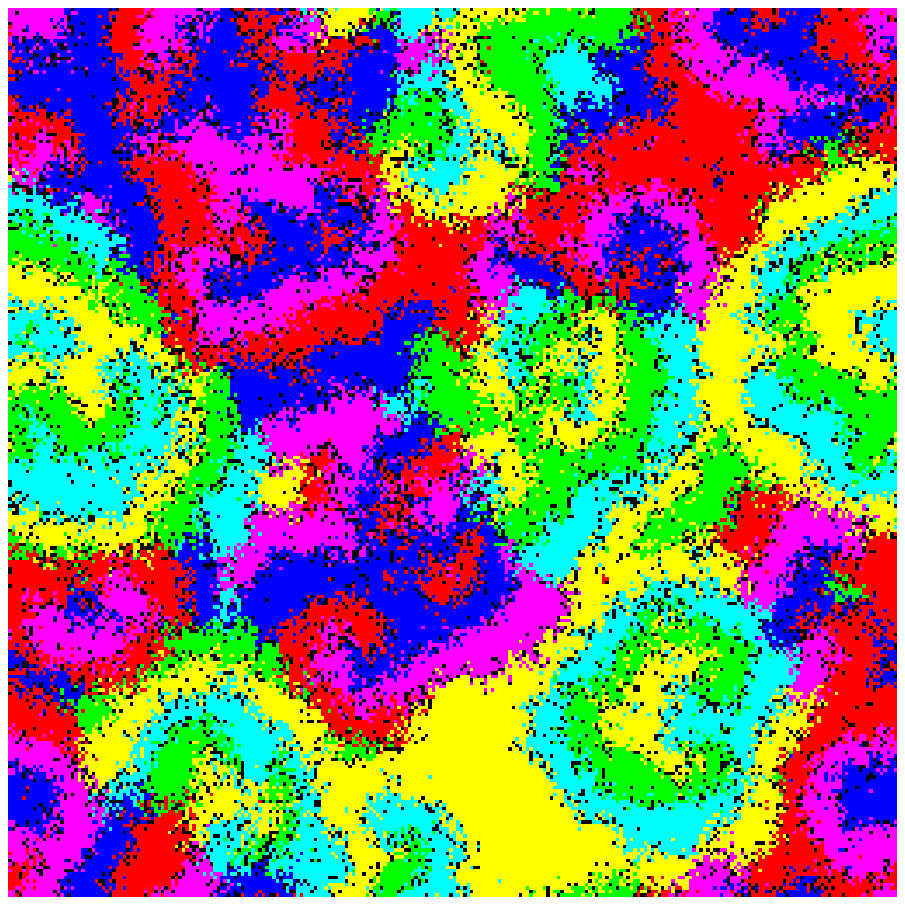}\\
  t=1000
\endminipage\hfill
\minipage{0.30\textwidth}%
  \centering
  \includegraphics[width=\linewidth]{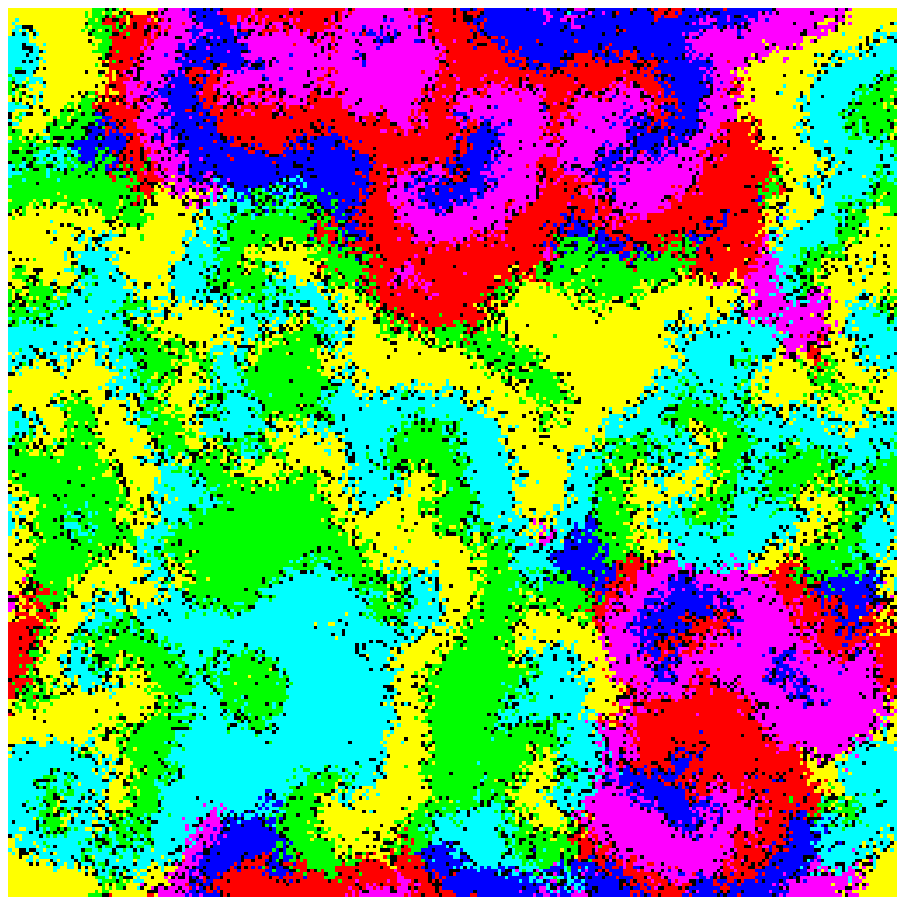}\\
  t=10000
\endminipage\\[0.5cm]
\minipage{0.30\textwidth}
  \centering
 \includegraphics[width=\linewidth]{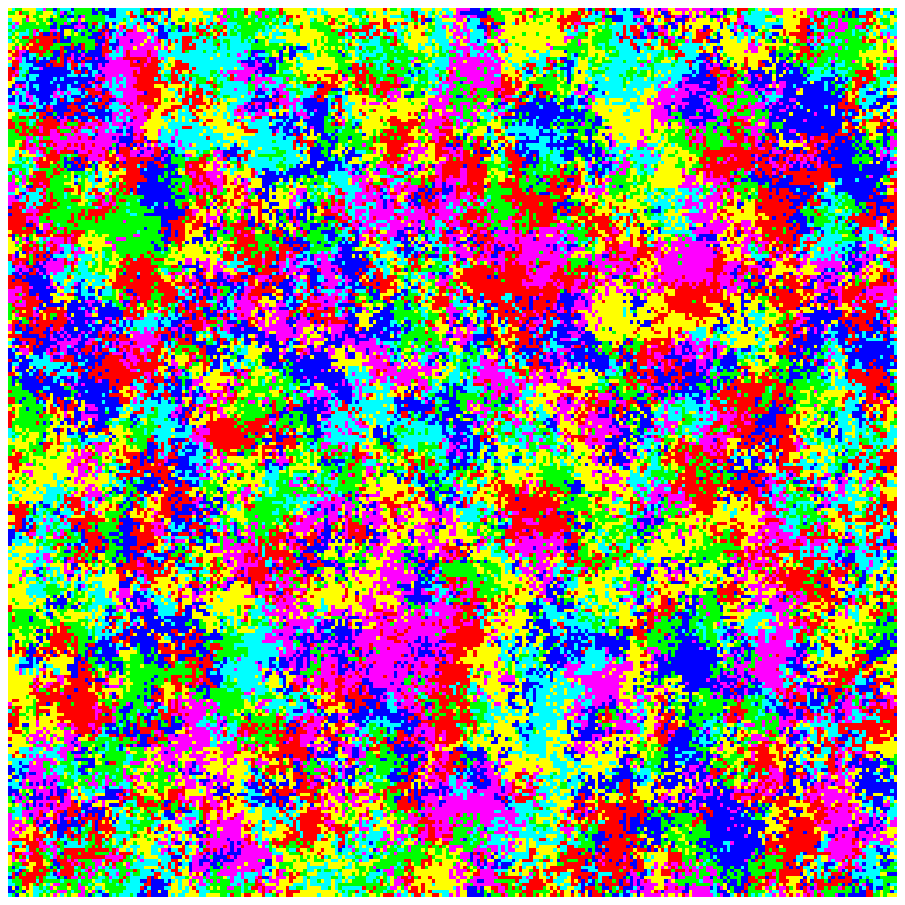}\\
  t=100
\endminipage\hfill
\minipage{0.30\textwidth}
  \centering
  \includegraphics[width=\linewidth]{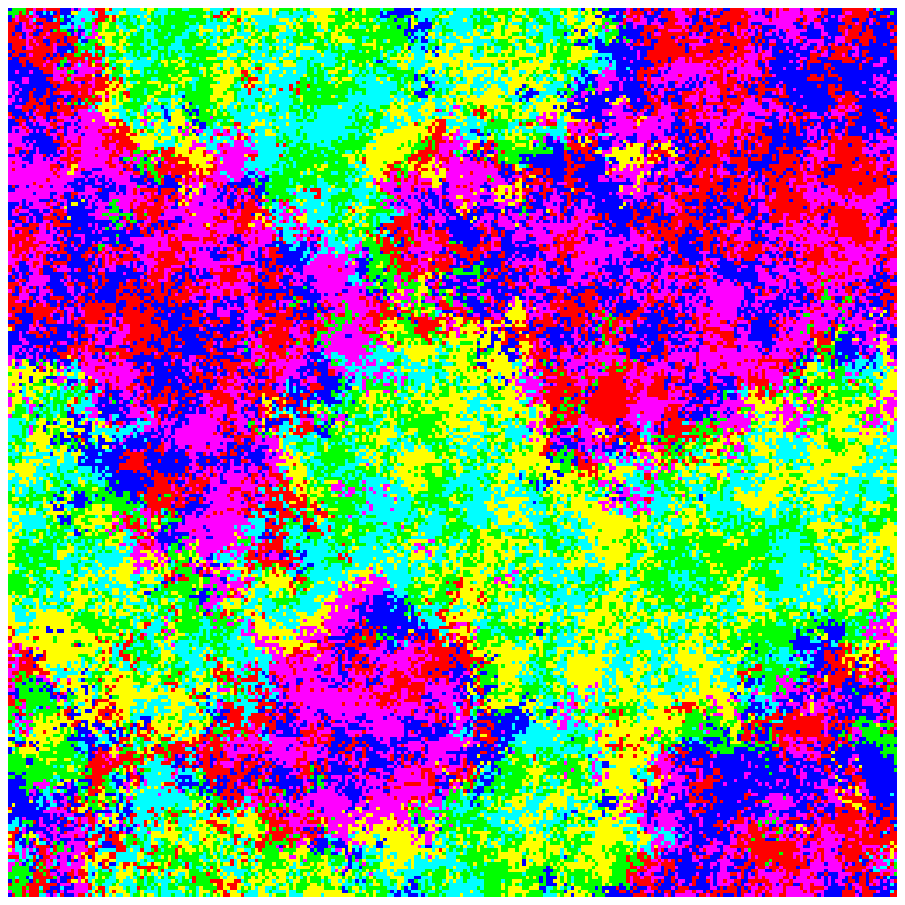}\\
  t=1000
\endminipage\hfill
\minipage{0.30\textwidth}%
  \centering
  \includegraphics[width=\linewidth]{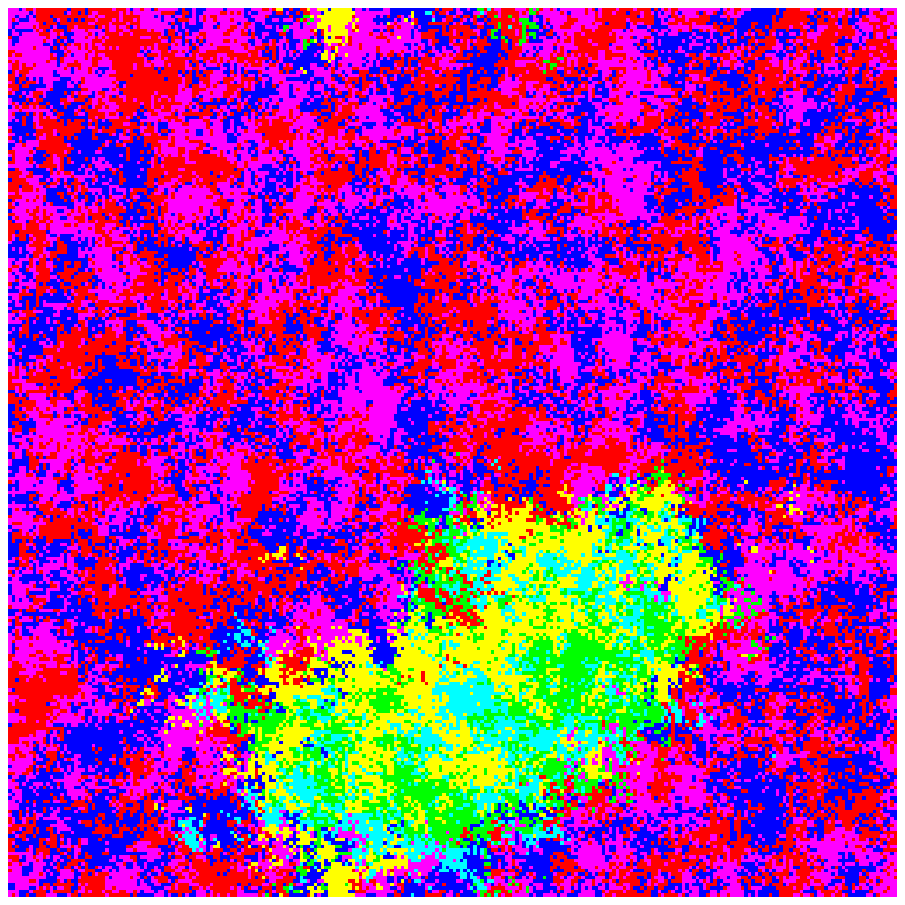}\\
  t=10000\\
\endminipage
\caption{(Color online) First row: Snapshots of the (6,3) system with empty sites at various times on a
$256\times256$ lattice showing the generation of the two types of competing domains where in each domain
the teams play a rock-paper-scissors game. Empty sites are indicated by black dots. Second row:
Snapshots of the (6,3) system without empty sites.}
\label{fig2}
\end{figure}

The spatial $(6,3)$ system provides one instance of intriguing emergent space-time patterns. This is illustrated
in the first row of Fig. \ref{fig2} through three different snapshots taken at different times since preparing the system
in a disordered initial state where each species has the same probability to occupy a lattice site. 
One observes the formation and coarsening of two different types of domains, each domain being
occupied by a team of three species (the bold arrows in Fig. \ref{fig1} indicate the two teams). Most interestingly,
every team inside a domain develops a cyclic three-species game, i.e. a (3,1) game, which results in the formation of spirals 
confined within the domains. It is this presence of large-scale structures
and their effects on the relaxation process that we address in the following.

We also simulated the Lotka-Volterra version of this system, where in the absence of empty sites predation and reproduction
take place simultaneously through the reaction
\begin{equation}
s_{i} + s_{j} \xrightarrow[]{\kappa} s_{i} + s_{i}~.
\label{eq:2}
\end{equation}
The only way for particles to move in that situation is through the swapping of particles located
on neighboring sites. As shown in the second row of Fig. \ref{fig2} we also have coarsening domains
in that case. However, the absence of empty sites does not
permit the formation of spirals, but instead every domain is occupied by patches containing individuals of one of the three species
forming an alliance. In absence of empty sites, the interfaces are rather fuzzy as
small clusters can enter into an enemy domain and survive for some time in this hostile environment.

\begin{figure}
\minipage{0.30\textwidth}
  \centering
  \includegraphics[width=\linewidth]{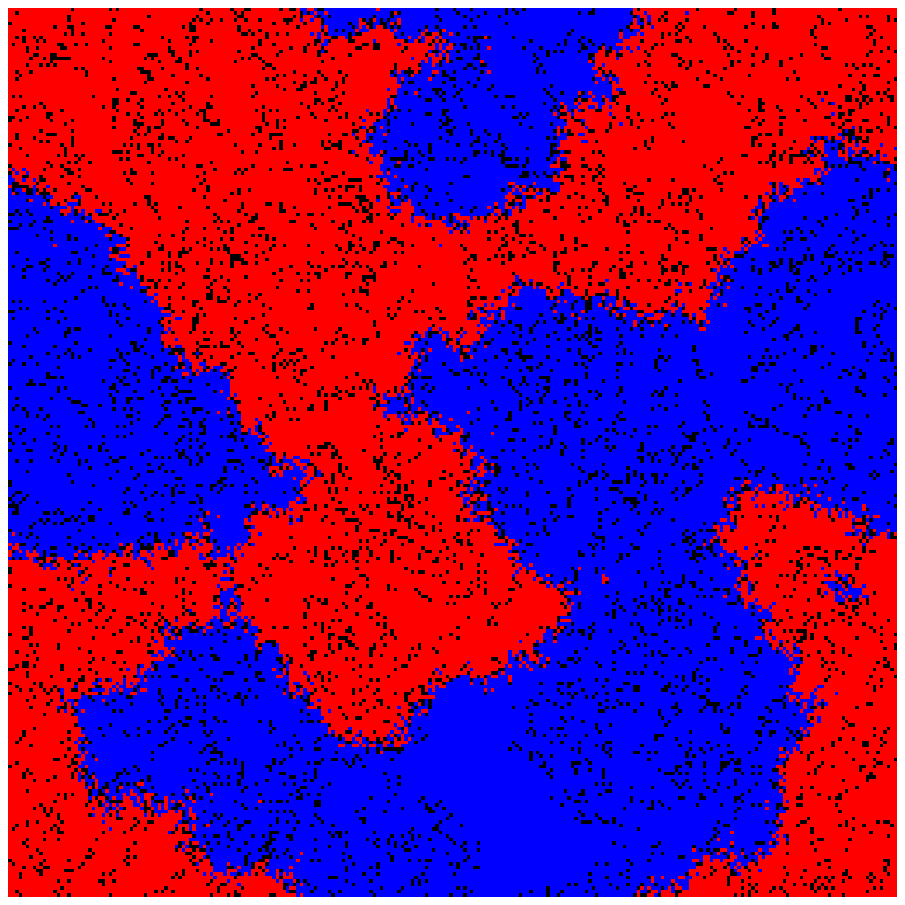}\\
  (a)
\endminipage\hfill
\minipage{0.30\textwidth}
  \centering
  \includegraphics[width=\linewidth]{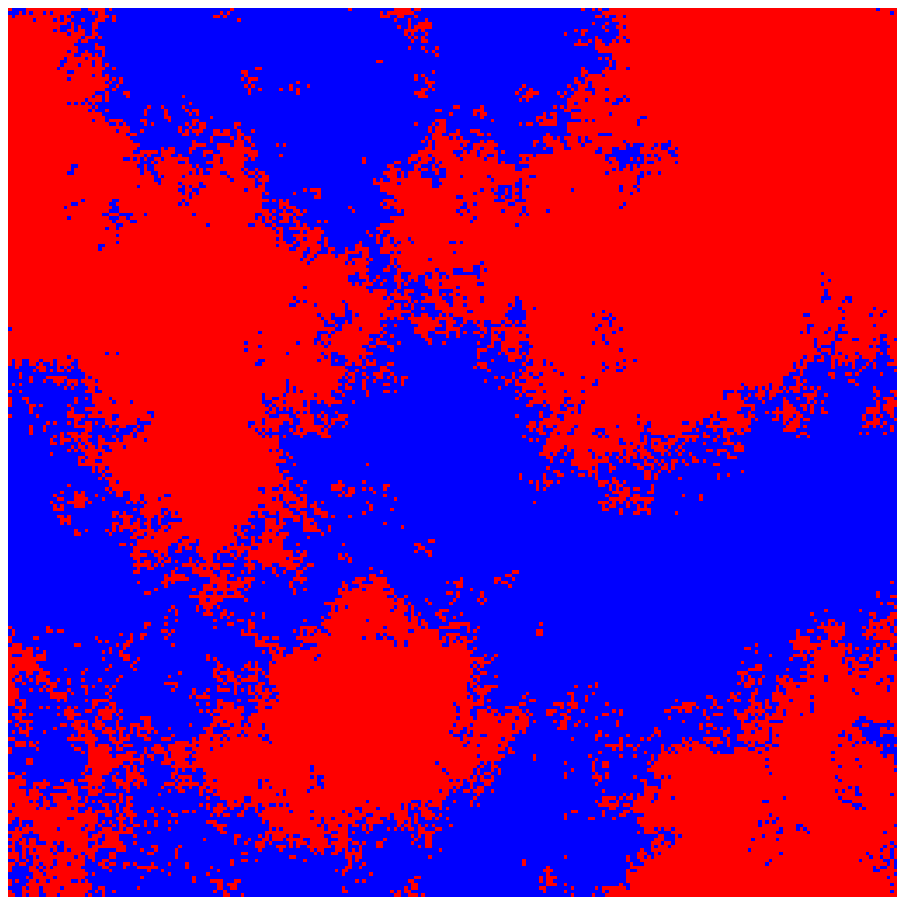}\\
  (b)
\endminipage\hfill
\minipage{0.30\textwidth}%
  \centering
  \includegraphics[width=\linewidth]{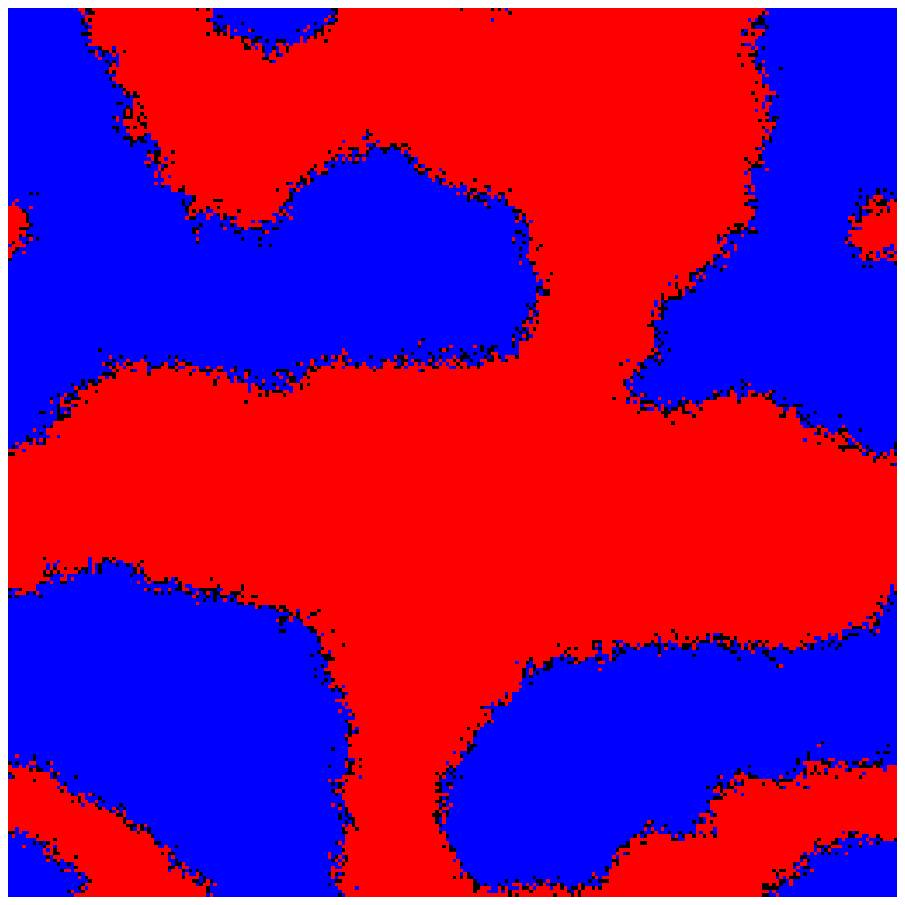}\\
  (c)
\endminipage
\caption{(Color online) (a) Snapshot of the (6,3) system with empty sites on a
$256\times256$ lattice at time $t=1,000$. All individuals of one team are shown in red,
whereas the individuals of the other team are colored in blue. Empty sites are indicated by black dots.
(b) The same for the (6,3) system without empty sites. For this case the boundaries between domains
are very diffuse. (c) Snapshot of the two-species (2,1) game with empty sites
on a $256\times256$ lattice at time $t=1,000$.}
\label{fig3}
\end{figure}

Before a more quantitative discussion, let us first have another look at the typical configurations shown in Fig.
\ref{fig2}. As the (6,3) system evolves in time, and this is true whether empty sites are present or not, 
the species separate into groups forming domains. Each domain
contains a team of three species, either (1,3,5) or (2,4,6), with cyclic interactions. In the May-Leonard version
with empty sites (top row) this rock-paper-scissors game yields spirals, whereas in the Lotka-Volterra version without empty sites
(bottom row) periodically changing patches form. Meanwhile interactions
between the two different teams only take place at or close to 
the domain boundaries. It is tempting to first neglect the internal dynamics
and only focus on the boundaries between domains. For this we "paint" in the same color, blue
or red, all individuals of one team, see Fig. \ref{fig3}a and Fig. \ref{fig3}b.  
We can compare these two snapshots to a snapshot in Fig. \ref{fig3}c of a (2,1) system with empty sites where two species prey on each other,
with the reaction scheme:
\begin{equation}
\begin{split}
s_{1} + s_{2} &\xrightarrow[]{\kappa} s_{1} + \emptyset \\
s_{2} + s_{1} &\xrightarrow[]{\kappa} s_{2} + \emptyset \\
s_{1} + \emptyset &\xrightarrow[]{\kappa} s_{1} + s_{1} \\
s_{2} + \emptyset &\xrightarrow[]{\kappa} s_{2} + s_{2} \\
s_{1} + s_{2} &\xrightarrow[]{\sigma} s_{2}+s_{1}\\
s_{2} + s_{1} &\xrightarrow[]{\sigma} s_{1}+s_{2}
\end{split}
\label{eq:3}
\end{equation}
where the first term represents the individual on the randomly selected site and the second the individual on the selected neighboring site.
In Fig. \ref{fig3} empty sites are again indicated by black dots. We note that in Fig.\ \ref{fig3}a we have empty sites both at the domain
boundaries and within the domains where they result from the effective dynamics within a team, whereas for the (2,1)
game empty sites show up only at the boundaries between domains. We also note that the interfaces for the (2,1) model are very sharp.
These colored snapshots, albeit very interesting, do not allow to make strong statements
about the time-dependent properties of domains and their boundaries. For this we need to perform a quantitative study using various space-time 
quantities.

\section{Space-time properties}

Much of our understanding of curvature driven coarsening results from in-depth studies of model systems like the
two-dimensional Ising model. When quenching this system to a temperature
below the critical temperature, two equilibrium states (one positively magnetized and the other negatively magnetized)
compete with each other, yielding the formation of a mosaic of domains that coarsen
over time. As shown in Fig. \ref{fig3}, a similar picture seems to hold for the (6,3) model when identifying 
as one of the states the pattern emerging from the interactions between the different members forming one of the 
two teams. In some recent studies of related many-species predator-prey models \cite{Avelino12a,Avelino12b,Avelino14a,Avelino14b,Avelino16}
the relevant length scale was found to increase as a square root of time, similar to what is found in the coarsening
regime of the Ising model.
These studies, however, focused (with one exception on which we comment below) on cases without the
formation of large-scale dynamic structures inside the domains.
Whereas the studies in \cite{Avelino12a,Avelino12b,Avelino14a,Avelino14b,Avelino16} only measured the density of empty sites and derived
from this quantity the typical length under the assumption that they are inversely proportional, we will in the following investigate
a wide range of quantities that have been extensively tested in the past for the Ising and related models. It follows from our results
that in the presence of non-trivial internal dynamics the values of the exponents governing
coarsening, aging, and interface fluctuations differ from those expected from a system with curvature driven dynamics.

\subsection{Space-time correlation and dynamical lengths}

We start our discussion with the space and time-dependent correlation function. As we will see, this quantity contains information
describing the structures inside of the growing domains as well as the domains themselves. 

\begin{figure} [h]
\includegraphics[width=0.60\columnwidth,clip=true]{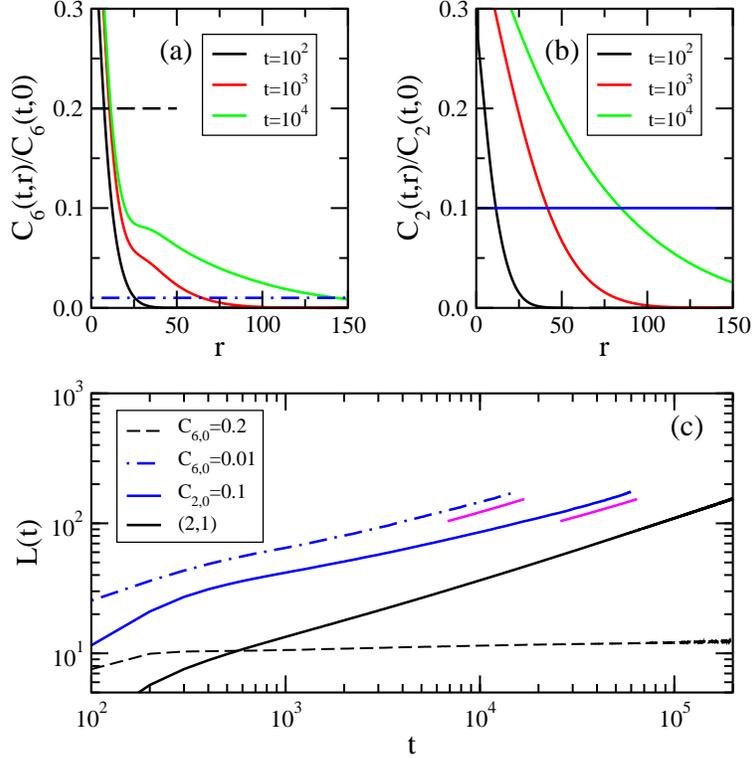}
\caption{\label{fig4} (Color online) 
Space and time-dependent correlation function for the (6,3) game with empty sites in two space dimensions when
(a) all six species are treated as separate and (b) species which make up a team are treated
as one. (c) Time-dependent correlation lengths extracted from the space-time correlation. The values
of $C_{6,0}$ and $C_{2,0}$ used for this are indicated by the horizontal line segments in (a) and (b), with the
color and line type matching those in (c). The solid magenta lines in (c) indicate the slope 0.43.
Also shown is the correlation length for the (2,1) system, with the slope 0.5 at later times as
expected for curvature driven coarsening. The data have been obtained for a system with $700 \times
700$ sites and result from averaging over 7,000 independent realizations of the noise.
}
\end{figure}

We measure the space and time-dependent correlation in two different ways: by treating all six
species as separate, see Fig. \ref{fig4}a, or by considering the species which make up a team as one,
see  Fig. \ref{fig4}b. In each case the space-time correlation is given by the quantity
\begin{equation}
C(t, r) = \sum\limits_i \left[ \langle n_i(\vec{r},t) n_i(0,t) \rangle -
\langle n_i(\vec{r},t)\rangle \, \langle  n_i(0,t) \rangle \right]~,
\end{equation}
where $r = \left| \vec{r} \right|$. The occupation number $n_i(\vec{r},t)$ is equal to 1 if at time $t$ an individual
from species $i$ sits on site $\vec{r}$ and zero otherwise. If we consider all six species, then
$i = 1, \cdots, 6$, whereas $i=1, 2$ if we consider as one the three species which make up a team.
For the former case we denote the correlation by $C_6(t,r)$, whereas for the latter we use $C_2(t,r)$.

Fig. \ref{fig4}a and \ref{fig4}b show our results for the (6,3) model with empty sites.
Inspection of Fig. \ref{fig4}a reveals that if we treat all species as separate, then the space-time correlation has two
very distinct regimes, one being a short distance regime that can be associated with the structures inside
the domains, the other being a long distance regime connected to domain coarsening. 
If, however, all species in one team are considered as one, then only one regime is observed, see Fig. \ref{fig4}b.
A time-dependent length $L(t)$ can be extracted from the correlation function 
by determining the distance $r$ at which $C_k(t,r)/C_k(t,0)$ ($k$ being 6 or 2, depending on whether or not we consider
all species to be separate) takes on a specific value $C_{k,0}$:
\begin{equation}
C_k(t,L(t))/C_k(t,0) = C_{k,0}~,
\end{equation}
as indicated by the horizontal line segments in Fig. \ref{fig4}a and \ref{fig4}b. Results of this procedure
are shown in Fig. \ref{fig4}c. Choosing a relatively large value like $C_{6,0} =0.2$ in Fig. \ref{fig4}a,
we obtain a length, characteristic of the formation of spirals inside the domains, that only displays
a weak dependence on time and approaches a plateau (black dashed line). On the other hand a low value like 
$C_{6,0} =0.01$ (blue dot-dashed line) allows us to extract a length related to domain growth.
After an early time behavior this length is proportional to that obtained from Fig. \ref{fig4}b, see the
full blue line in Fig. \ref{fig4}c. For long times the slopes of both these lengths approach the value $x_C = 0.43(1)$,
which is smaller than the value 1/2 expected for purely curvature driven coarsening. For comparison we 
include in Fig. \ref{fig4}c the correlation length obtained in the (2,1) system (full black line) which 
does display a slope of 1/2.

While the difference between 0.43 and 1/2 might seem to be small, we find it to be very robust and not to change when
choosing a different value for $C_{2,0}$. Taken at face value, this result indicates that in the presence
of non-trivial internal dynamics the value of the exponent $x_C$ governing domain growth differs from that of curvature driven
coarsening. In the remaining text, further and stronger evidence that non-trivial internal dynamics alter relaxation processes
is presented through the analysis of other quantities. 

\subsection{Two-times autocorrelation function}

\begin{figure} [h]
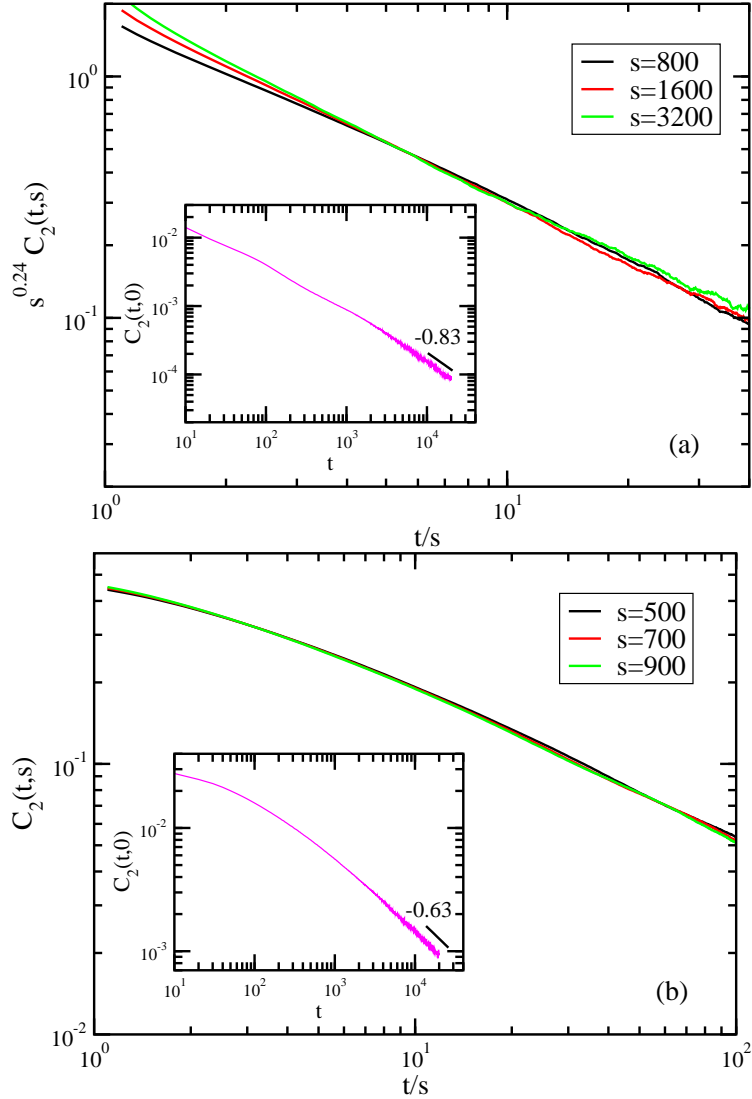

\includegraphics[width=0.60\columnwidth,clip=true]{figure5a.eps}\\
\includegraphics[width=0.60\columnwidth,clip=true]{figure5b.eps}
\caption{\label{fig5} (Color online)
Aging scaling of the two-times autocorrelation function $C_2(t,s)$ for different waiting times $s$: 
(a) (6,3) system with empty sites and (b) (2,1) system.
The insets show the time-dependence of the autocorrelation for $s=0$. The data for (6,3), obtained
for systems with $700 \times 700$ sites, result from averaging over 30,000 independent runs for $s>0$,
whereas for $s=0$ the average is taken over 100,000 realizations of the noise. For (2,1) we used
2,900 realizations for every value of $s$.
}
\end{figure}

In many systems relaxation processes are accompanied by dynamical scaling. This is especially true
for coarsening systems, as for example the two-dimensional Ising model quenched to temperatures below
the critical point \cite{Henkel10}. Aging scaling is best probed through two-times quantities like
the two-times autocorrelation function $C(t,s)$. In the case of a power-law growth of the typical domain
size, simple aging scaling of the form \cite{Henkel10}
\begin{equation}
\label{eq:aging}
C(t,s) = s^{-b} f(t/s)
\end{equation}
is expected, where the scaling function $f(y)$ displays a power-law $f(y) \sim y^{- \lambda}$ when
$y \gg 1$. The scaling form (\ref{eq:aging}), which is expected to hold when the waiting time $s$, the observation 
time $t$ as well as their difference $t-s$ are large compared to any microscopic timescale, has been observed
in many different systems as for example spin glasses and magnets \cite{Henkel10}.

In our systems aging scaling can be probed through the two-times autocorrelation function
\begin{equation}
C_2(t,s) = \sum\limits_{i=1}^2 \left[ \langle n_i(0,t) n_i(0,s) \rangle -
\langle n_i(0,t)\rangle \, \langle  n_i(0,s) \rangle \right]~,
\end{equation}
where we consider all species of one team as identical. The autocorrelation $C_6(t,s)$, where all species are
considered to be separate, is not suited for this purpose. Indeed, $C_6(t,s)$ is very sensitive to the dynamics
within the domains, yielding a periodic pattern in the presence of spirals.

As verified in Fig. \ref{fig5}a, aging scaling is indeed observed for the (6,3) system with empty sites, with
exponents $b=0.24(1)$ and $\lambda = 0.83(1)$. Interestingly, these values differ markedly from the values
$b=0$ and $\lambda = 0.63$ of the two-dimensional Ising model undergoing phase ordering. In Fig. \ref{fig5}b we
probe whether the scaling (\ref{eq:aging}) is also encountered in the (2,1) case and find a behavior
compatible with that of the Ising model quenched below the critical point. We point out that the (6,3)
model without empty sites also shows for large waiting times the same aging scaling as the (2,1) system,
with exponents $b=0$ and $\lambda = 0.63$ (not shown here).

\subsection{Density of empty sites}

\begin{figure} [h]
\includegraphics[width=0.60\columnwidth,clip=true]{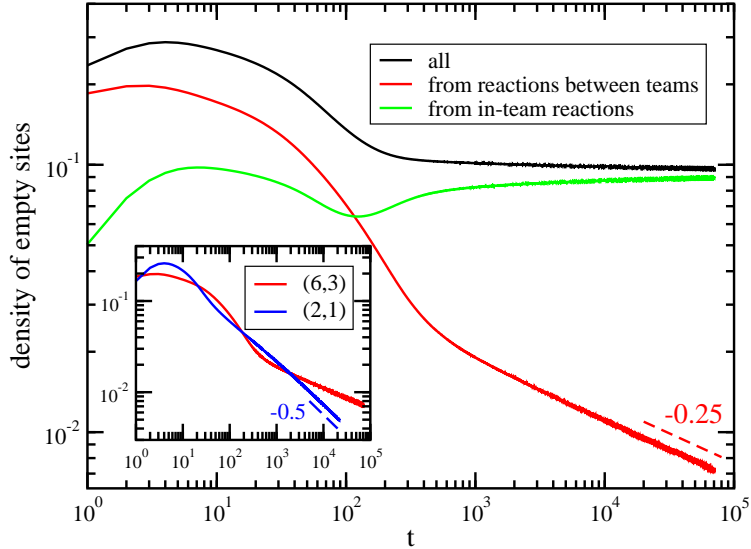}
\caption{\label{fig6} (Color online)
Time evolution of the density of empty sites in the (6,3) model with the reaction scheme (\ref{eq:1}).
After some transient behavior the density of the empty sites created by reactions between members from one team
(green line) approaches a constant value. On the other hand, the density of empty sites that result from interactions between the
teams (red line) decays algebraically with an exponent $x_E = -0.25(1)$, as indicated by the dashed red line. The data, obtained
for a system with $700 \times 700$ sites, result from
averaging over 7,000 different runs. Inset: comparison of the (6,3) density of empty sites resulting from 
interactions between teams with the density of empty sites obtained for the (2,1) model.
The dashed blue line indicates a decay with an exponent $-0.5$.
}
\end{figure}

Following previous work by Avelino {\it et al.} \cite{Avelino12a,Avelino12b,Avelino14a,Avelino14b,Avelino16}
we have also investigated the time-dependence of the density of empty sites. Empty sites are created in reactions
involving a predator and its prey, see the reaction scheme (\ref{eq:1}). In cases with domain
coarsening a large number of empty sites are formed at the boundaries between the domains. This yields a network of
strings of empty sites that provides an easy way to follow domain growth and coarsening over time.
Focusing on cases without production of empty sites inside the domains (either because the domains are pure
due to phase segregation or composed exclusively of neutral partners), Avelino {\it et al.} argue that during the
coarsening regime the characteristic
length should vary inversely proportional to the number of empty sites. It follows that for curvature driven 
coarsening the number of empty sites should vanish as $t^{x_E}$ with $x_E=-1/2$. Analyzing a range of different models,
they find values of $x$ close but slightly smaller than $-1/2$. In the inset of Fig. \ref{fig6} we verify this for the (2,1) model,
arguably the simplest model with coarsening of pure domains, and find indeed the value $x_E = -0.495(10)$.

The situation is more complicated for the (6,3) reaction scheme where empty sites are also created inside of the 
domains, due to the rock-paper-scissors game between members of the same team. Consequently, we need to distinguish
between the two different types of empty sites. For this every empty site created through a predator-prey interaction
is labeled as either due to an in-team interaction or due to an interaction between the two teams, depending on the species
involved in the interaction. As shown in Fig. \ref{fig6}, after some transient early time behavior, related to the formation
of spirals within the domains, the density of empty sites produced in in-team reactions (green line) approaches a plateau, as expected
for the appearance of stable spiral patterns. In contrast to this, the density of empty sites produced in reactions involving
individuals from both teams (red line) decays algebraically with an exponent $x_E = -0.25(1)$. While an algebraic decay with time is expected for
empty sites formed at the boundaries between domains, the value we find is markedly different from the value $-1/2$
expected for simple curvature driven coarsening and found by us for the (2,1) model, see the inset of Fig. \ref{fig6}.
The collisions of spirals at the domain boundaries strongly slow down the elimination of empty sites that are originally formed
through interactions between the different teams. 
We also remark that the postulated simple relationship \cite{Avelino12a,Avelino12b,Avelino14a,Avelino14b,Avelino16}
between the exponent $x_C$ of the correlation length and the
exponent $x_E$ of the density of empty site, $x_E = - 1/x_C$, 
does not hold for the (6,3) model with non-trivial internal dynamics.

In \cite{Avelino12a} the density of empty sites was also investigated for a slightly different version of our (6,3) model (model V
in that paper). The authors did not provide any figure with the corresponding data, but merely quote the value $x_E = -0.429 \pm 0.029$
for the exponent describing the decrease of the number of empty sites as a function of time. As they consider 
``only the empty spaces which have as some of the four immediate neighbors individuals from the 2 groups: {1, 3, 5} and {2, 4, 6},''
we ran simulations in which the counting conforms to this criterion and found that also this subset of empty sites
decays with the same exponent $-0.25(1)$ as in Fig. \ref{fig6}. We therefore cannot reproduce these results
from that earlier study. 

\subsection{Interface width}

The properties of an interface are best studied by
preparing the system in the following initial state. Consider a system with $L \times H$ sites and 
separate the system into two equal parts of width $L/2$ each. 
Each half is then occupied by individuals randomly selected among the
species forming one of the teams (in cases with empty sites, we also leave a certain fraction of sites initially unoccupied).
In this way we have an initial state with a straight interface that separates the system into two halves, each half being occupied 
exclusively by members of one of the two teams.
During the updates particles located at the left (right) edge of the system can only interact 
with three neighbors, namely their north, south, and right (left) neighbors.

\begin{figure}
\minipage{0.30\textwidth}
  \centering
  \includegraphics[width=\linewidth]{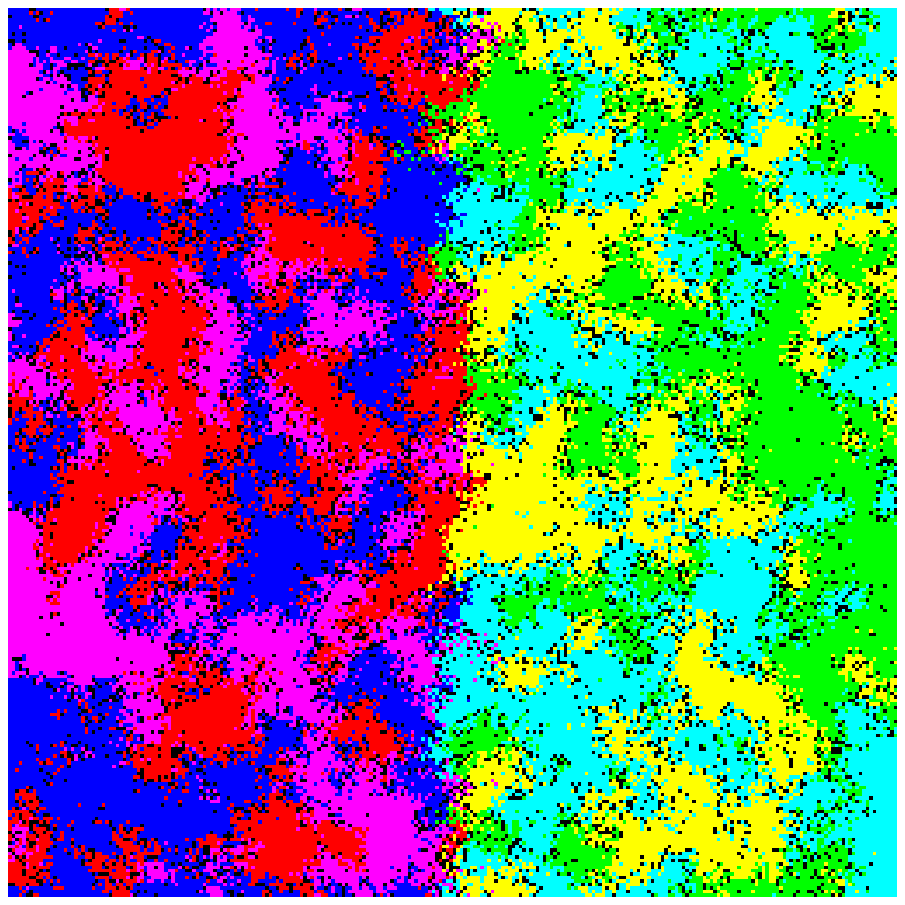}\\
  t=100
\endminipage\hfill
\minipage{0.30\textwidth}
  \centering
  \includegraphics[width=\linewidth]{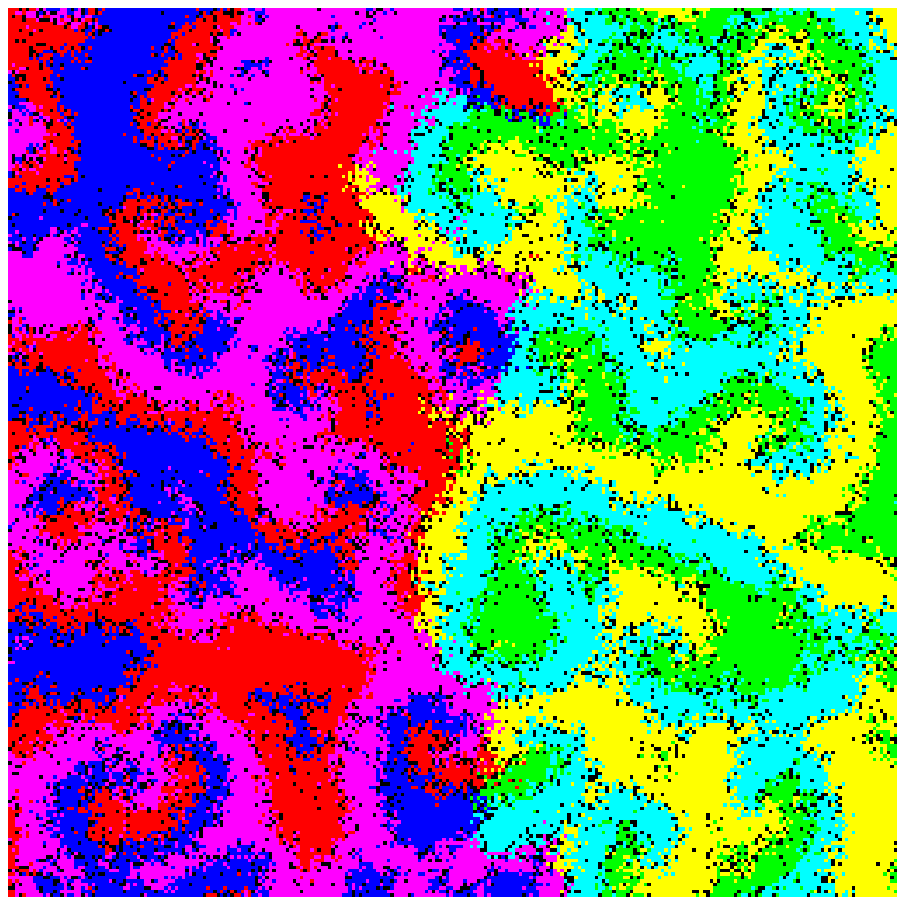}\\
  t=1000
\endminipage\hfill
\minipage{0.30\textwidth}%
  \centering
  \includegraphics[width=\linewidth]{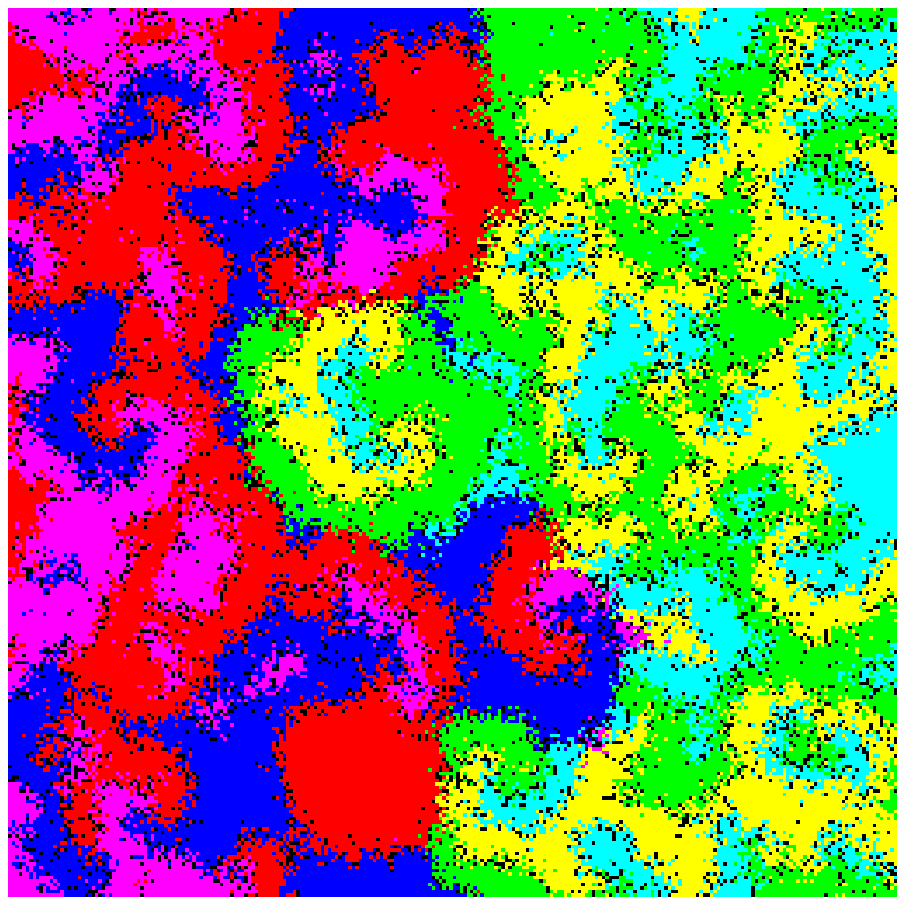}\\
  t=10000
\endminipage\\[0.5cm]
\minipage{0.30\textwidth}
  \centering
 \includegraphics[width=\linewidth]{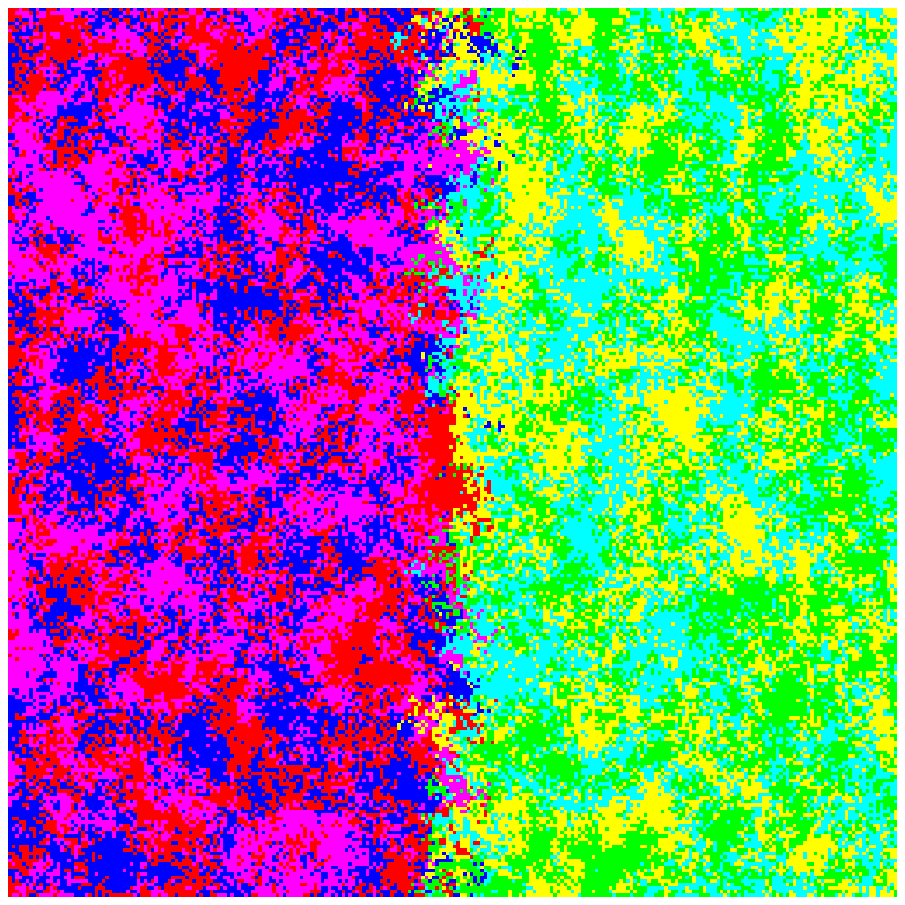}\\
  t=100
\endminipage\hfill
\minipage{0.30\textwidth}
  \centering
  \includegraphics[width=\linewidth]{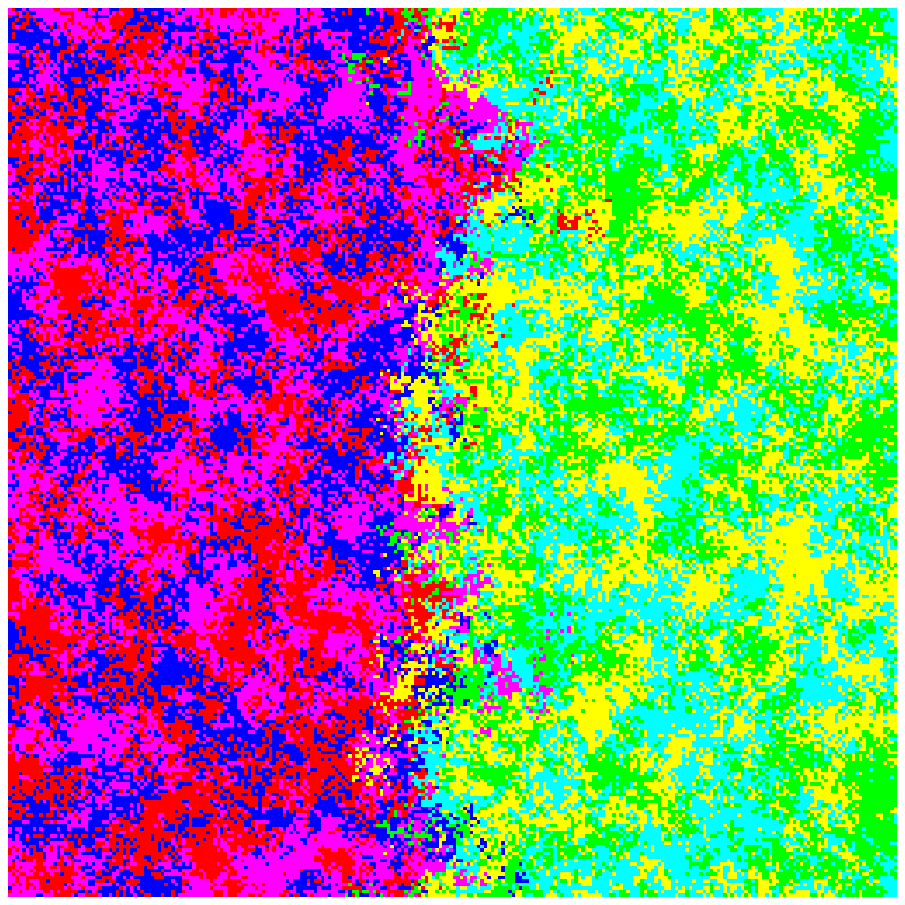}\\
  t=1000
\endminipage\hfill
\minipage{0.30\textwidth}%
  \centering
  \includegraphics[width=\linewidth]{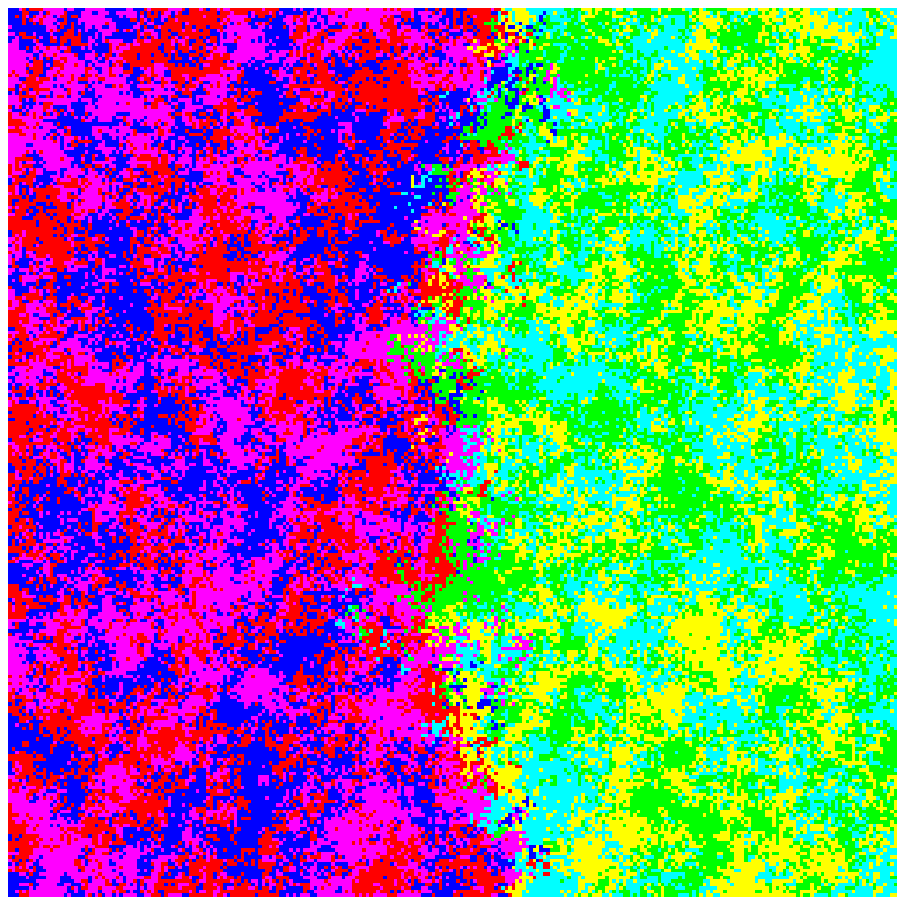}\\
  t=10000\\
\endminipage\\[0.5cm]
\minipage{0.30\textwidth}
  \centering
 \includegraphics[width=\linewidth]{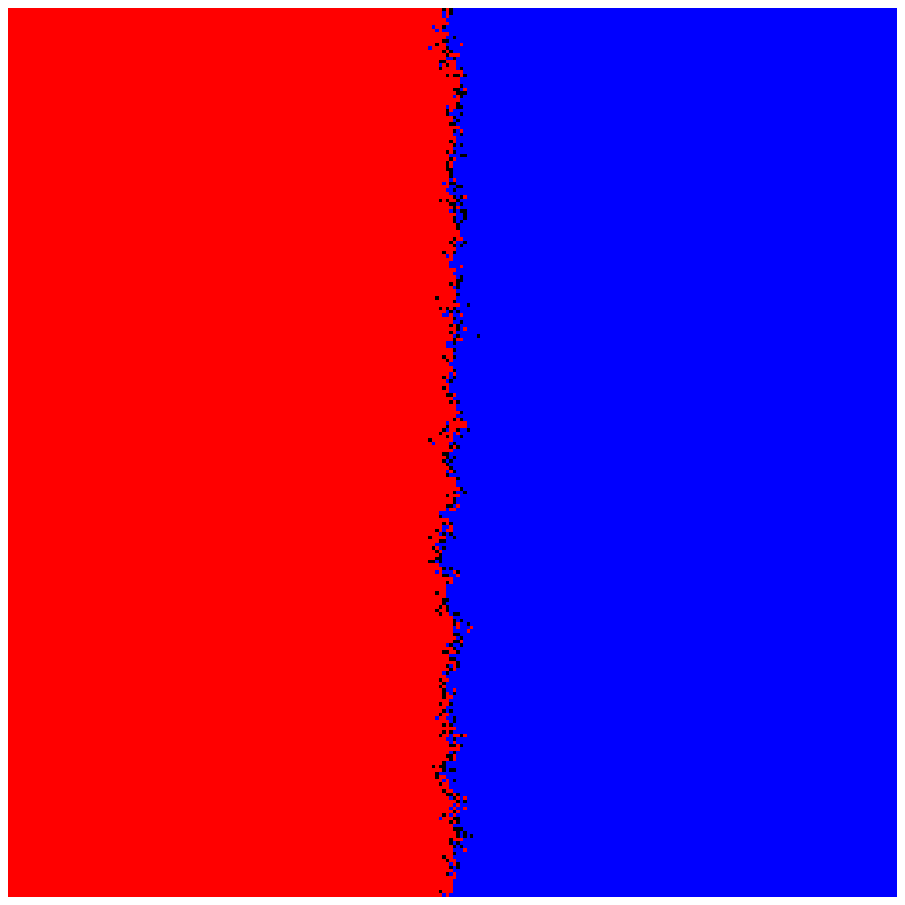}\\
  t=100
\endminipage\hfill
\minipage{0.30\textwidth}
  \centering
  \includegraphics[width=\linewidth]{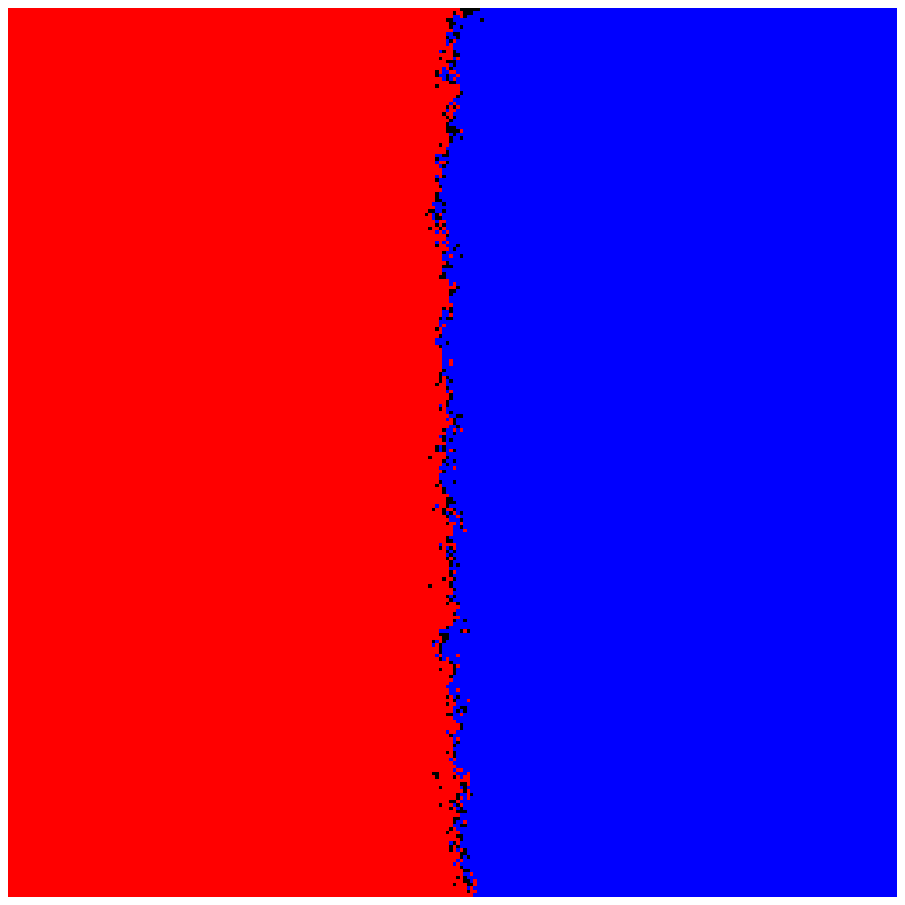}\\
  t=1000
\endminipage\hfill
\minipage{0.30\textwidth}%
  \centering
  \includegraphics[width=\linewidth]{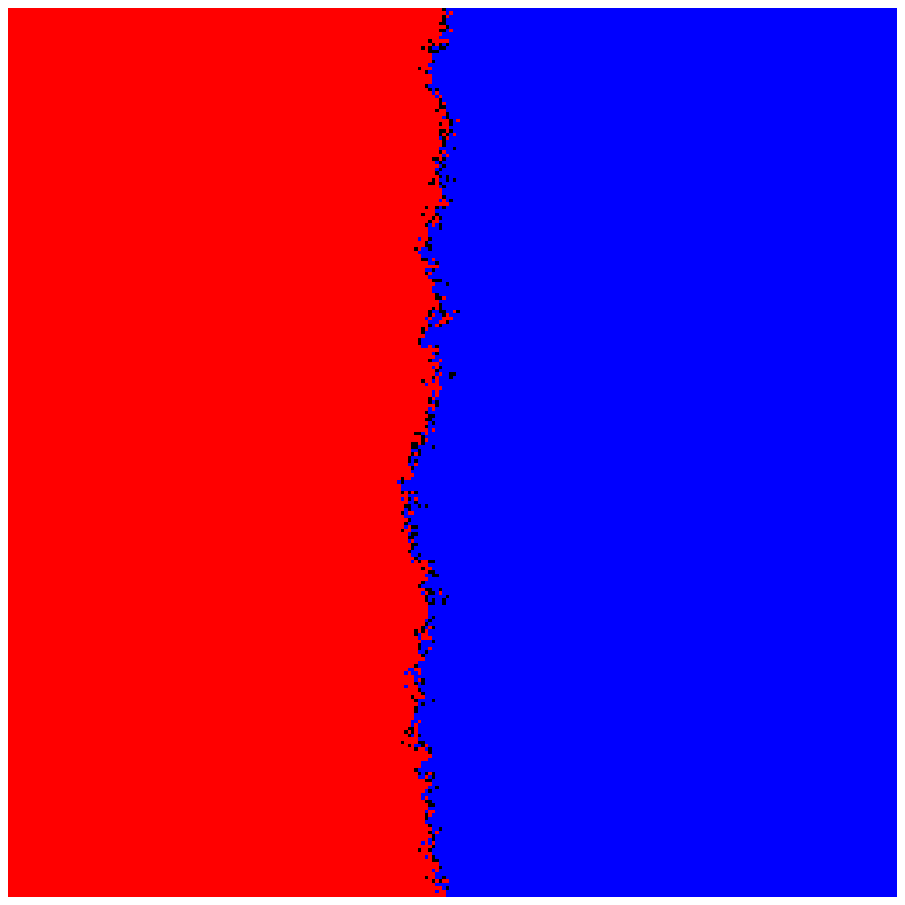}\\
  t=10000\\
\endminipage
\caption{(Color online) 
First row: Snapshots at various times of an interface in the (6,3) system with empty sites on a
$256\times256$ lattice. At $t=0$ the system is separated in two halves where all the sites in one of the halves 
are randomly occupied by individuals from one team only. After formation of the spirals large interface
fluctuations are observed. Second row: The same, but now for the (6,3) system without empty sites. Due to the absence of 
large structures in each half, the interface fluctuations are much less pronounced. Third row: The same, but
now for the (2,1) system with each species initially occupying one half of the system.
}
\label{fig7}
\end{figure}

The snapshots shown in Fig. \ref{fig7} 
indicate that the presence of spirals has a major impact on
the properties of the interface. Indeed, for the (6,3) model with empty sites, see the first row,
the successive wavefronts initiate large-scale coherent
fluctuations of the interface which strongly contrasts with both the (6,3) case without empty sites (second row)
and the (2,1) model (third row) which display much more localized fluctuations of the interface.

These differences in the roughening of the interface can be quantified through an investigation of
the interface width. In order to do so we first need to determine the local position of the interface
which can be rather diffuse, due to the interactions and exchanges
taking place at the boundary between the two teams. We introduce a variable $S_{i,j}$ that characterizes
the occupation of the site $(i,j)$ \cite{Roman12} at some time $t$. If site $(i,j)$ is occupied, then $S_{i,j} =\pm 1$,
depending on whether the individual at that site belongs to team 1 or team 2. If the site is unoccupied,
then $S_{i,j} = 0$. We then follow 
\cite{Virgilis05,Muller96} and determine for each row $j$ the value $l$ that minimizes the sum
\begin{equation}
u(l)= \sum\limits_{i=1}^L \left[ S_{i,j} - s \left( i-l \right) \right]^2~,
\end{equation}
where $s(v)$ is the Heaviside step function, with $s = 1$ for $v < 0$ and $s=-1$ for $v > 0$. 
From these local positions $l(j)$ we can then determine the mean position of the interface at time $t$:
\begin{equation}
\overline{l} = \frac{1}{H} \sum\limits_{j=1}^H l(j)~
\end{equation}
as well as the interface width $W(t)$ given by the standard expression
\begin{equation}
W(t) = \sqrt{ \frac{1}{H} \sum\limits_{j=1}^H \left( l(j) - \overline{l} \right)^2 }~.
\end{equation}

\begin{figure} [h]
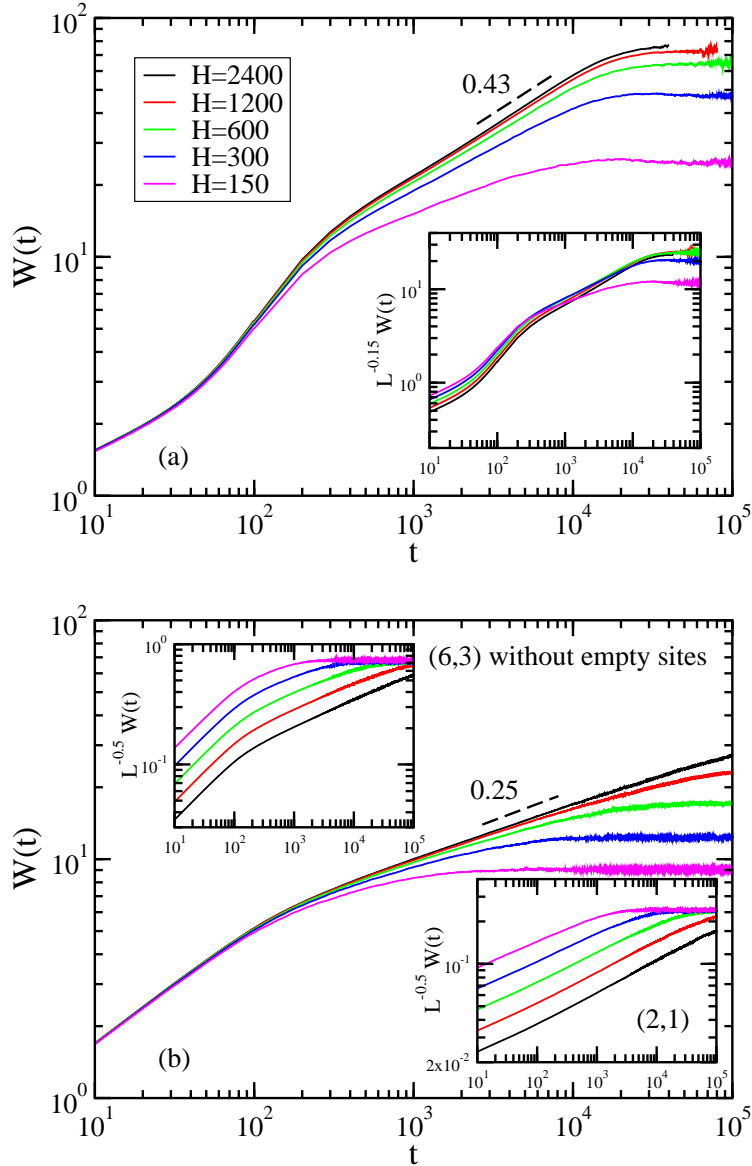

\includegraphics[width=0.60\columnwidth,clip=true]{figure8a.eps}\\[0.5cm]
\includegraphics[width=0.60\columnwidth,clip=true]{figure8b.eps}
\caption{\label{fig8} (Color online) Interface width $W(t)$ for the different systems. (a) (6,3) model with empty
sites. The large scale interface fluctuations induced by the spirals, see first row of snapshots in Fig. \ref{fig7},
yield values for the growth exponent $\beta = 0.43(1)$ and roughness exponent $\alpha = 0.15(2)$ that differ from the 
standard Edwards-Wilkinson values $\beta=1/4$ and $\alpha = 1/2$. (b) (6,3) model without empty sites for which
the interface width displays the Edwards-Wilkinson scaling, see also inset in the upper left corner. The inset
in the lower right corner plots $L^{-1/2} W$ vs time for the (2,1) case for which we again find the
Edwards-Wilkinson exponents. The system sizes are $500 \times H$ where the different values of $H$ are given
in the legend of panel (a). The data result from averaging over at least 8,000 independent runs.
}
\end{figure}

As shown in Fig. \ref{fig8} for rectangular systems of $500 \times H$ sites with $H$ ranging from 150 to 2400,
the interface width for all three cases exhibits after an early time behavior the expected two regimes: a correlated
regime where the width increases algebraically with time: $W \sim t^\beta$, with the growth exponent $\beta$,
followed by a regime where the fluctuations, and therefore the width, saturate at a value that depends on $H$:
$W \sim H^\alpha$, with the roughening exponent $\alpha$. A previous study \cite{Roman12} revealed that for the (4,1) game, where two teams
composed of neutral partners are formed, these two exponents take on the values $\beta = 1/4$ and $\alpha = 1/2$ of the
Edwards-Wilkinson universality class \cite{Edwards82}. As shown in Fig. \ref{fig8}b, we find the same result both
for the (6,3) game without empty sites as well as for the (2,1) game. This is not surprising as even for the
(6,3) game without empty sites the fluctuations, that ensue when patches of individuals from one species come close
to the interface, are short-distance fluctuations that have no persistent impact on the scaling properties of the interface. This is different for the 
(6,3) system with empty sites, and therefore with non-trivial internal dynamics, where the wave fronts due to the spiral patterns
result in large-scale fluctuations and an enhanced roughening of the interface. Indeed, see Fig. \ref{fig8}a, the accelerated
roughening of the interface due to the spirals yields a growth exponent $\beta = 0.43(1)$, much larger than the Edwards-Wilkinson
value. In the saturation regime, we find, after discarding systems too small to allow the formation of
well-formed spirals, a good scaling of the saturation width
with the roughening exponent $\alpha = 0.15(2)$. These values of the two exponents, which do not agree with those 
expected for any of the standard
universality classes for interface fluctuations, unambiguously reveal the decisive impact non-trivial dynamics 
inside coarsening domains can have on the properties of the domain boundaries.

\section{Conclusion} 
Far from equilibrium intriguing space-time patterns can emerge from very simple microscopic rules. Coarsening domains,
encountered in a large variety of situations, provide well-known examples. Usually, the dynamics within the domains
is rather trivial (for the Ising model quenched below the critical point the spins deep inside a domain behave essentially
like spins in equilibrium). In this paper we have presented results for a case where the dynamics within the domains
is non-trivial and takes the form of spirals due to the cyclic competition of three different species. As our work shows,
these spirals have an impact both on the coarsening process and on the interface fluctuations, yielding values of the standard
exponents very different from those expected for curvature driven coarsening. 

Whereas we focused on the case where all rates are identical, it
is an interesting question whether (and if so, to what extent) the values
of the different exponents depend on the details of the model as for example
the values of the predation and swapping rates. While on general grounds
a high degree of universality could be expected (this is supported by preliminary
data for cases where the predation and swapping rates are
no longer identical), it might be interesting to check this explicitly through
additional studies focusing, for example, on cyclic situations where a species
attacks their different preys with different rates.

We have obtained our results through large-scale Monte-Carlo simulations, and it remains a challenge to come up
with a coarse-grained description that would allow a more analytical investigation of the space and time-dependent
properties of coarsening domains that contain emergent spiral patterns.

The six-species model exhibiting these intriguing
properties is only one example of patterns within patterns that emerge in a larger family of models introduced in the context 
of population dynamics. When considering nine species one can have three different types of domains, similar to the three-states
Potts model at low temperature, where within each domain a rock-paper-scissors game takes place. One also can have the appearance 
of smaller spirals inside larger ones in the case of an hierarchical game \cite{Brown17}. We plan to investigate these and other
cases in detail in the future.

\begin{acknowledgments}
This work is supported by the US National
Science Foundation through grant DMR-1606814.
\end{acknowledgments}

\end{document}